\documentclass[aps,pra,showpacs,superscriptaddress,10pt]{revtex4-2}

\usepackage[utf8]{inputenc}
\usepackage[english]{babel}

\usepackage{amssymb}
\usepackage{color}
\usepackage{amsmath}
\usepackage{graphicx}
\usepackage{bm}
\usepackage{braket}
\usepackage{booktabs}
\usepackage{multirow}
\usepackage{url}
\usepackage{tikz}
\usepackage{quantikz}

\begin{document}

\title{Quantum Non-Moduler Multiplication with QFT-Based Multi Input Parallelized Adder}

\author{Murat Kurt}
\email{kmuratphysics@gmail.com}
\affiliation{Department of Software Engineering, Samsun University, 55420 Samsun, T\"{u}rkiye}

\author{Sel\c{c}uk \c{C}akmak}
\email{selcuk.cakmak@samsun.edu.tr}
\affiliation{Department of Software Engineering, Samsun University, 55420 Samsun, T\"{u}rkiye}

\author{Azmi Gen\c{c}ten}
\email{gencten@omu.edu.tr}
\affiliation{Department of Physics, Ondokuz May{\i}s University, 55139 Samsun, T\"{u}rkiye}

\begin{abstract}
In this study, we propose an efficient quantum multiplication approach based on a QFT-assisted parallelized addition scheme. The multiplication stage is implemented using a structure composed entirely of Toffoli gates, which generate partial products. In the second stage, these partial results are accumulated using a QFT-based adder. Unlike conventional QFT-based arithmetic circuits, the proposed design eliminates the repeated application of QFT and inverse QFT (IQFT) operations during intermediate summation processes. This leads to a significant reduction in the total gate count and circuit complexity, enabling a more resource-efficient implementation. To demonstrate the feasibility of the proposed approach, a quantum circuit that performs the multiplication of two 3-bit numbers is designed. The circuit is tested and validated using IBM quantum simulators. The results indicate that the proposed method provides a more efficient alternative to traditional quantum multiplication techniques in terms of gate cost and circuit depth.

\vspace{10pt}
\begin{description}
\item[Keywords] Quantum arithmetic, Quantum multiplication, QFT adder, Circuit optimization

\end{description}
\end{abstract}

\maketitle

\section{Introduction}
Quantum computing encodes information into the quantum states and processes it according to the principles of quantum mechanics. The fundamental properties of quantum states, such as superposition, entanglement and the no-cloning, provide quantum computing with advantages over classical computation \cite{bennett2000}. This advantage is commonly referred to as quantum supremacy. The property of superposition enables parallel information processing, potentially leading to exponential speedups in algorithmic performance. Entanglement and the no-cloning property, on the other hand, allow secure and consistent transmission of information.

The fundamental unit of information in quantum computing is the qubit. A qubit is a two-level ($d = 2$) quantum system and can be represented as $\ket{\psi}=\alpha\ket{0}+\beta\ket{1}$. This expression indicates that a qubit can exist in a superposition of two basis states. In special cases where either $\alpha$ or $\beta$ is zero, the qubit collapses to a definite state, namely $\ket{0}$ or $\ket{1}$, which are referred to as pure states. Once information is encoded into qubits, quantum circuits are designed to implement quantum algorithms. Although there is no strict classification of quantum algorithms, they can generally be categorized into quantum search algorithms, Quantum Fourier Transform (QFT)-based algorithms, the HHL algorithm for solving linear systems of equations, quantum eigenvalue solvers, and Hamiltonian simulation algorithms. In addition, quantum arithmetic algorithms, both QFT-based and non-QFT-based, which are directly related to the focus of this study, have attracted significant attention in recent years.\par

Arithmetic operations used in classical computing have been adapted to quantum computing frameworks and implemented as quantum circuits \cite{vedral1995}. In particular, several quantum circuit designs have been proposed for addition \cite{Gossett1998,cuccaro2004}. The first QFT-based addition circuit was introduced by Draper \cite{draper2000}. Subsequent studies have analyzed both QFT-based and non-QFT-based circuits for addition and multiplication \cite{beauregard2003,florio2004}. Over time, the scope of quantum arithmetic operations has expanded to include subtraction, division, arithmetic averaging, comparison, multiplication by a constant, and weighted summation \cite{ruizperez2017,sahin2020}. Furthermore, with the development of high-dimensional quantum systems, arithmetic operations originally implemented using qubits have been extended to qudit-based systems, demonstrating a reduction in the required number of quantum gates \cite{wang2020,pavlidis2017}. \par

Within QFT-based arithmetic operations, one of the earliest and most fundamental approaches for multiplication is the elemantery school multiplication algorithm. In this method, pairwise multiplications between the bits of the input numbers are performed, and the resulting partial products are accumulated to obtain the final result. The corresponding quantum circuit generally consists of two main components: the multiplication stage and the accumulation stage \cite{ramezani2023}. Alternatively, the Karatsuba algorithm, which divides numbers into smaller parts and performs recursive multiplications, has been proposed. However, this decomposition increases the circuit cost in a quantum setting. While the elementary school multiplication algorithm has a time complexity of $O(n^2)$, the Karatsuba algorithm achieves $O(n^{\log_2 3})$ \cite{karatsuba1962}. The Toom–Cook algorithm further reduces this complexity to $O(n^{\log_3 5})$\cite{toom1963}. The Schönhage–Strassen algorithm, which represents numbers as polynomials and performs multiplication via convolution, achieves a time complexity of $O(n \log n \log \log n)$ \cite{schonhage1971}. In addition, iterative multiplication algorithms and optimized methods for large integers have also been developed \cite{furer2007,Harvey2021}. Moreover, quantum circuit designs for both modular and non-modular multiplication, as well as exponentiation-based multiplication, have been proposed \cite{cho2020,zhan2023,chuang2018}. Time complexity in such circuits depends on parameters such as the number of inputs, gate count, and circuit depth. However, due to uncertainties in physical constraints such as gate execution times and qubit coherence times, the exact circuit cost remains an open challenge. \par

In this study, we propose a novel quantum circuit that performs non-modular multiplication based on the elementary school multiplication algorithm. In the multiplication stage, partial products are generated using a sequence of Toffoli gates. These intermediate results are then accumulated using a QFT-based parallelized addition circuit designed for summing $N$ $n$-bit numbers \cite{cakmak2023}. The proposed circuit is generalized for two $n$-bit inputs, and the number of required ancillary qubits is expressed as a function of $n$. Finally, the proposed design is implemented for the multiplication of two 3-bit numbers and validated using the IBM quantum simulator.

\section{Theory}

In this section, the Toffoli gate which is one of the fundamental gates in quantum computing is discussed. Subsequently, the processor implementing the Quantum Fourier Transform (QFT) is examined both at the functional level and in terms of its quantum circuit representation. Finally, a quantum circuit that performs QFT-based addition is presented and the elementary school multiplication method will be recalled.

\subsection{Toffoli Gate}

The Toffoli gate, a fundamental three-qubit quantum logic gate, operates with two control qubits and one target qubit. Also known as the controlled-controlled-NOT (CCNOT) gate, it performs a NOT operation on the target qubit if and only if both control qubits are in the $\ket{1}$ state. Otherwise, the state of the target qubit remains unchanged. The circuit representation of the Toffoli gate is illustrated in Fig.~1.

\begin{figure}[ht]
    \centering
    \includegraphics[width=0.25\linewidth]{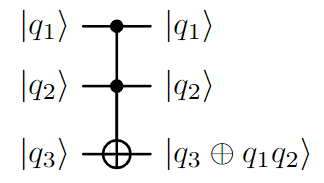}
    \caption{Quantum Circuit Representation of the Toffoli Gate}
\end{figure}

\subsection{Quantum Fourier Transform}

In quantum computing, the Quantum Fourier Transform (QFT) enables the parallel processing of multiple inputs by transforming input states into superposition. The QFT circuit consists of Hadamard gates, controlled phase shift gates, and SWAP gates. This transformation is a fundamental component in many pioneering quantum algorithms \cite{camps2021,jozsa1998,barenco1996,cao2011}. The mathematical formulation of the QFT is given in Eq.~(1). In this expression, $\ket{a}$ represents an $n$-qubit quantum state, while $\ket{k}$ denotes the output obtained after applying the QFT operator. The resulting state is a superposition. In simulations performed using quantum computing libraries such as Qiskit, quantum states are expressed through linear algebraic representations. In this context, both $\ket{a}$ and $\ket{k}$ are represented as $2^n \times 1$ dimensional vectors, while the QFT operator itself is defined as a $2^n \times 2^n$ unitary matrix. A generalized QFT circuit is illustrated in Fig.~2.

\begin{equation}
\mathrm{{QFT\ket{a}} = \frac{1}{\sqrt{2^n}}\sum_{k=0}^{{2^n}-1}e^{2\pi i.ak/{2^n}}\ket{k}}
\end{equation}

\begin{figure*}[!htb]
    \centering
    \resizebox{0.90\linewidth}{!}{
        \includegraphics{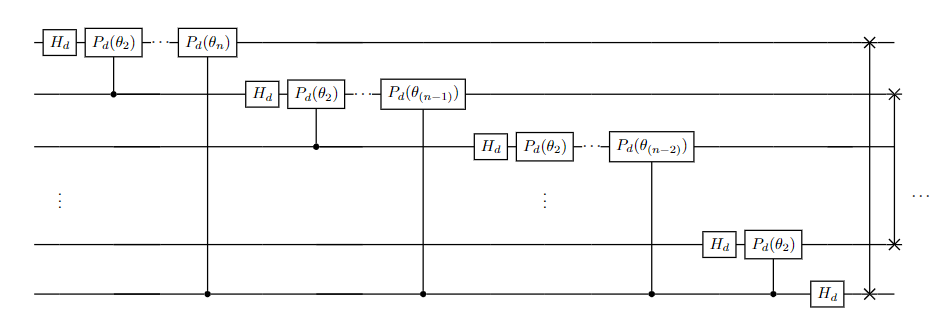}
    }
    \caption{Generalized QFT Circuit (For qubit systems d = 2)}
\end{figure*}

The Hadamard gate acts on a single qubit and produces a superposition state. The matrix representation and the operation of the Hadamard gate are given in the following equations.

\begin{equation}
    Hadamard =\frac{1}{\sqrt{2}}\begin{bmatrix}
        1 & 1 \\
        1 & -1 \\
    \end{bmatrix}
\end{equation}

\begin{equation}
    H\ket{q}=\frac{1}{\sqrt{2}}(\ket{0}+e^{2 \pi i(0.q)} \ket{1})
\end{equation}\par

The phase shift gate, similar to the Hadamard gate, acts on a single qubit. If the input is $\ket{0}$, it leaves the state unchanged, whereas for the input $\ket{1}$, it introduces a phase factor. The matrix representation of the phase shift gate for qubits is given below.

\begin{equation}
    P(\theta_k) =\begin{bmatrix}
        1 & 0 \\
        0 & e^{i\theta_k} \\
    \end{bmatrix}
\end{equation}

Here, $\theta_k = \frac{2\pi}{2^k}$. However, the QFT operator involves controlled phase shift gates. When the control qubit is in the $\ket{1}$ state, the phase shift gate given in Equation~4 is applied to the target qubit. The following equation provides the matrix representation of the phase shift gate for qubits \cite{pavlidis2017}.

\begin{equation}
    CP(\theta_k)= \sum_{j=0}^{1} \sum_{m=0}^{1} {e^{\frac{i2\pi}{2^k}jm}} \ket{j}\bra{j} \otimes \ket{m}\bra{m}
\end{equation}\par

\subsection{QFT-Based Parallelized Addition Circuit}

QFT-based quantum circuits for both modular and non-modular addition have been proposed. In these circuits, to compute the sum $a_1 + a_2$, the input numbers are first represented in the binary number system. Each bit of these binary representations corresponds to an input of the circuit. After applying the QFT to the input state $\ket{a_1}$ representing the first number, the state $\ket{\phi(a_1)}$ is obtained. As stated, $\ket{\phi(a_1)}$ is a superposition state. Subsequently, a subcircuit composed of controlled phase shift gates, referred to as the adder, is applied such that the inputs representing $a_2$ act as control qubits, while $\ket{\phi(a_1)}$ serves as the target. The resulting state of this operation is $\ket{\phi(a_1 + a_2)}$. 

In the final step, the inverse QFT (IQFT) operator is applied, yielding the result of $a_1 + a_2$ in the binary number system. However, when the number of operands exceeds two, this approach becomes computationally expensive. For instance, in the case of summing $N$ numbers such as $a_1 + a_2 + a_3 + \dots + a_N$, the QFT and IQFT operations must be applied at each step. This leads to an increase in the required number of gates, as well as in circuit depth and time complexity. To mitigate this overhead, the QFT and IQFT operations are applied only once at the beginning and the end of the circuit, respectively. The QFT-based parallelized quantum addition circuit that performs the summation of $N$ $n$-bit numbers is illustrated in Figure~3. In the circuit, $t$ additional qubits are used to store the carry bits generated during the addition process.

\begin{figure*}[!htb]
    \centering
    \resizebox{0.8\linewidth}{!}{
        \includegraphics{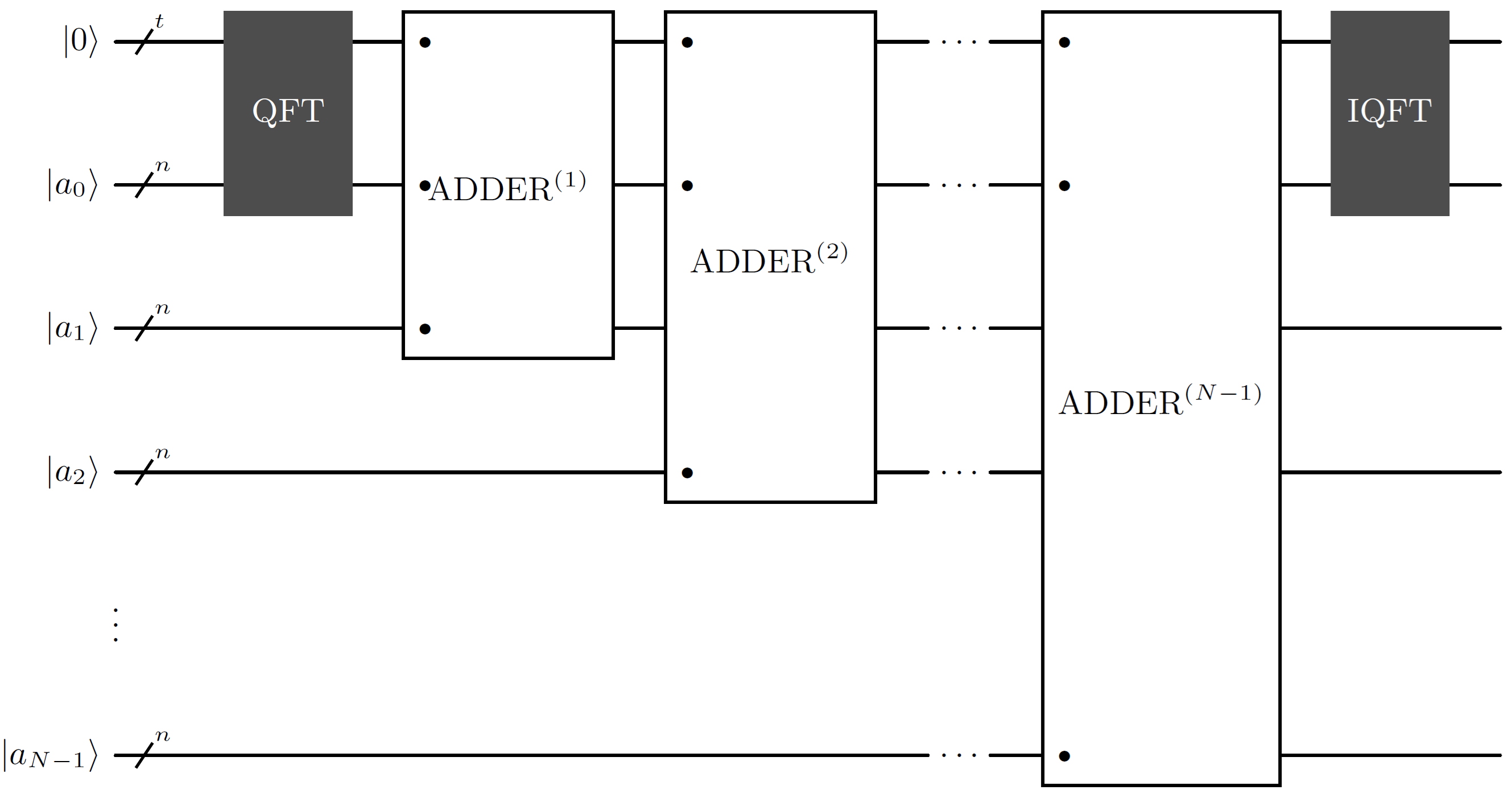}
    }
    \caption{QFT-Based Parallelized Addition Circuit}
\end{figure*}

\subsection{Elementary School Multiplication Method}

To adapt classical multiplication algorithms to quantum computing, it is first necessary to understand their fundamental structure. Specifically, the schematic representation of the multiplication of two 3-bit numbers $x_1x_2x_3$ and $y_1y_2y_3$ is shown below in operational sceheme and Figure 4. This process consists of two main parts. The first part involves performing bitwise multiplication operations. In the first step, the bit $y_3$ is multiplied sequentially with $x_3$, $x_2$, and $x_1$. The resulting bits from each multiplication are right-aligned and written in the section referred to as the second part, forming the first partial product.

In the second step, the bit $y_2$ is multiplied sequentially with $x_3$, $x_2$, and $x_1$. The resulting bits are written as the second partial product in the second part, shifted one bit to the left relative to the first partial product. The empty position created by this shift is filled with zero. Finally, the bit $y_1$ is multiplied sequentially with $x_3$, $x_2$, and $x_1$, and the resulting bits are written with an additional one-bit left shift compared to the previous row. The two-bit empty space on the right-hand side is filled with zeros. From the resulting arrangement, it is observed that three partial products, each consisting of five bits, are obtained. In constructing the quantum multiplication circuit, it is required that all numbers to be added have the same bit length; therefore, the empty positions in the first two rows are also padded with zeros. Consequently, the final multiplication result is obtained by summing these three 5-bit numbers.

\[
\begin{array}{c@{}c@{}c@{}c@{}c@{}c@{}c@{}}
& & & x_1 & x_2 & x_3 \\
& & & y_1 & y_2 & y_3 \\
& & \times \\ 
\cline{3-6} \\
&\hspace{1em}\textcolor{red}{\textbf{0}} &\hspace{1em}\textcolor{red}{\textbf{0}} &\hspace{1em}y_3x_1&\hspace{1em}y_3x_2 &\hspace{1em}y_3x_3 \\ [1ex]
&\hspace{1em}\textcolor{red}{\textbf{0}} &\hspace{1em}y_2x_1&\hspace{1em}y_2x_2 &\hspace{1em}y_2x_3 &\hspace{1em}\textcolor{red}{\textbf{0}} \\[1ex]
&\hspace{2em}y_1x_1&\hspace{1em}y_1x_2 &\hspace{1em}y_1x_3 &\hspace{1em}\textcolor{red}{\textbf{0}} &\hspace{1em}\textcolor{red}{\textbf{0}} \\[1ex]
\hline \\
&s_0&s_1&s_2&s_3&s_4&s_5\\
\end{array}
\]

\begin{figure*}[!htb]
    \centering
    \resizebox{0.8\linewidth}{!}{
        \includegraphics{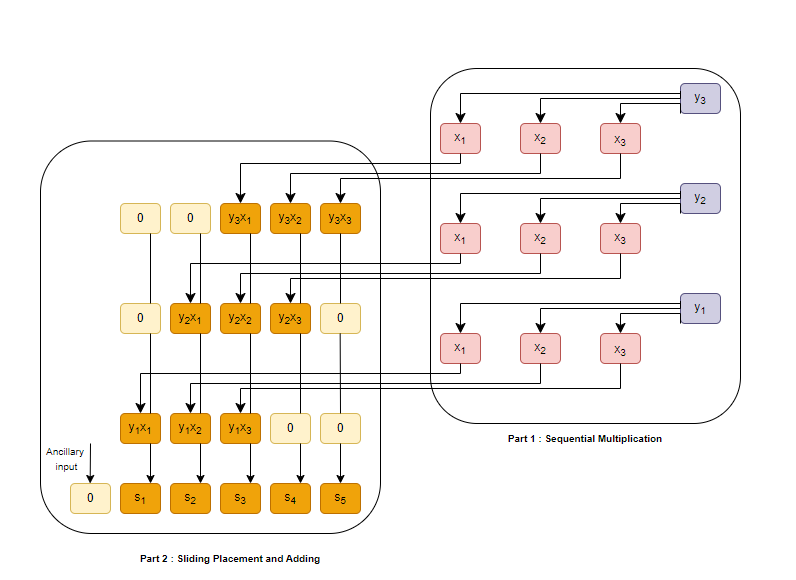}
    }
    \caption{Elementary Multiplication Operation Diagram}
\end{figure*}

As shown in Fig.~4, this process consists of two stages. In the first stage, the multiplication between bits is performed, while in the second stage, the results obtained from these multiplications are summed. In this way, the final result of the multiplication operation is obtained. In Fig.~5, the QFT-based quantum circuit that performs this addition process is illustrated.

\begin{figure*}[!htb]
    \centering
    \resizebox{0.8\linewidth}{!}{
        \includegraphics{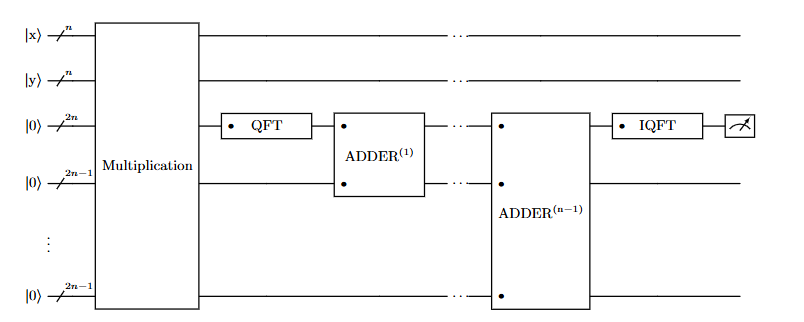}
    }
    \caption{Non-Modular Multiplication with QFT-Based Multi-Input Parallelized Adder Circuit ($\mathrm{n}$-bit $\mathrm{N}$-input)\cite{kurt2024}}
\end{figure*}

\section{Results and Discussion}

In this study, a more efficient approach is proposed for implementing classical multiplication within the framework of quantum computing. In the proposed method, the bitwise multiplication operations in the first stage of the classical multiplication algorithm are realized using Toffoli gates. The intermediate results obtained from these multiplications are then combined using a QFT-based parallelized addition scheme. This approach enables a reduction of $n-1$ QFT and IQFT operations out of the $n$ required for $n$-bit numbers. Figure~6 illustrates the generalized quantum circuit that performs the multiplication of two $n$-bit numbers. This circuit computes the product of two numbers encoded in the quantum states $\ket{x_1x_2 \dots x_n}$ and $\ket{y_1y_2 \dots y_n}$. For this purpose, a total of $2n$ qubits are used as inputs. In addition, $n$ auxiliary registers are employed to store the intermediate multiplication results corresponding to each bit of the multiplier, where each register initially consists of $2n-1$ qubits. However, as shown in the figure, the first of these auxiliary registers is extended by one additional qubit, increasing its size to $2n$. This extra qubit is required to store the carry information generated during the addition process. Consequently, the proposed circuit comprises a total of $2n^2 + n + 1$ qubits. The circuit consists of two main parts: a multiplication stage, entirely constructed from Toffoli gates, which performs bitwise multiplications, and an addition stage, which combines the intermediate results using a QFT-based addition scheme.

\subsection{Part of Multiplication}

This part of the circuit is composed entirely of Toffoli gates. For two $n$-bit numbers, a total of $n^2$ Toffoli gates are required. The generalized structure of this section is illustrated in Figure~5. 

In the first step, a Toffoli gate is applied such that the $n$-th bits of the numbers $y$ and $x$ act as control qubits, while the last qubit of the first $2n-1$ block serves as the target. Subsequently, while keeping the $n$-th bit of $y$ fixed as one of the control qubits, Toffoli gates are applied sequentially with the remaining bits of $x$ as control qubits. In each application, the target is shifted one line upward. In this manner, the $n$-th bit of $y$ is multiplied with each bit of $x$ in sequence.

In the second step, the $(n-1)$-th bit of $y$ is multiplied with all bits of $x$ using Toffoli gates. However, in this case, the first multiplication result is not written to the lowest qubit of the second $2n-1$ block, but rather to the qubit immediately above it. This shift corresponds to a one-bit left shift, as observed in classical multiplication. 

These operations are iteratively continued, starting from one line above at each step. The application of the Toffoli gate and the resulting output values can be computed mathematically as shown in the following equation.

\begin{equation}
    \begin{split}
    &\bigotimes_{i,j=1}^{n} \text{Toffoli} \ket{y_{n+1-i},x_{n+1-j},0} \\
    &= \bigotimes_{i,j=1}^{n} \ket{y_{n+1-i},x_{n+1-j}} \ket{0\oplus y_{n+1-i} x_{n+1-j}}
\end{split}
\end{equation}

\begin{figure*}[htbp]
    \centering
    \resizebox{0.9\linewidth}{!}{
        \includegraphics{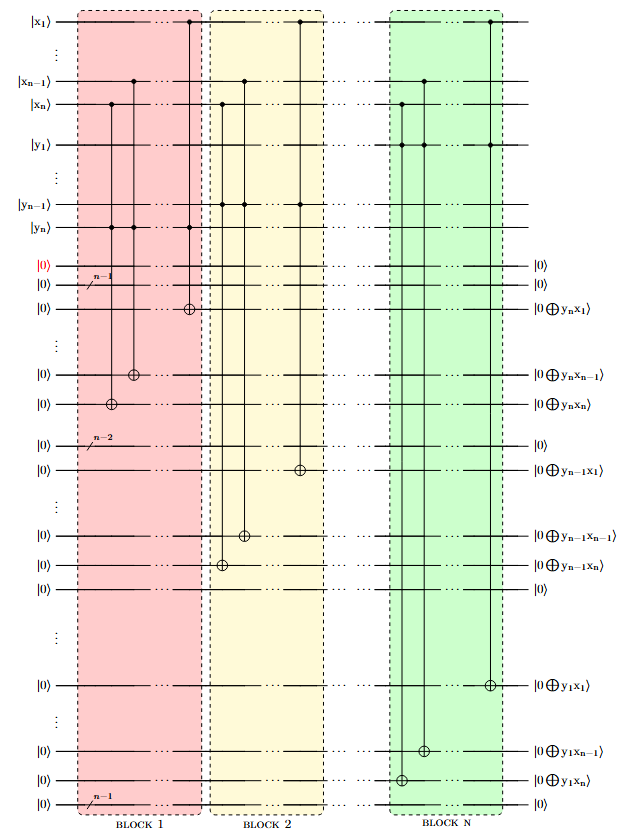}
    }
    \caption{General Multiplier Circuit}
\end{figure*}

\subsection{Three bit number Example}

In this section, an implementation of the generalized quantum multiplication circuit is presented for the case $n=3$. The input states are chosen as $\ket{x} = \ket{111}$ and $\ket{y} = \ket{101}$. As shown in Figure~7, the multiplication stage consists of nine Toffoli gates. 

The first group of three Toffoli gates performs the multiplication of $\ket{y_3}$ with $\ket{x_3}$, $\ket{x_2}$, and $\ket{x_1}$, respectively. The second group of three Toffoli gates computes the products of $\ket{y_2}$ with $\ket{x_3}$, $\ket{x_2}$, and $\ket{x_1}$. Finally, the last group of three Toffoli gates yields the products of $\ket{y_1}$ with $\ket{x_3}$, $\ket{x_2}$, and $\ket{x_1}$. 

The resulting partial products are arranged by starting from the lowest line and shifting each subsequent result one line upward. This procedure corresponds to the classical multiplication scheme, where each partial product is shifted one bit to the left with respect to the previous one. 

In the second stage, a QFT-based parallelized addition circuit is applied. By including one auxiliary qubit, the lowest six qubits—containing the first partial product—are transformed into the Fourier domain via the QFT. Subsequently, the addition stage, composed of controlled phase shift gates, is applied sequentially: first using the second group of five qubits as control qubits, and then using the third group of five qubits as control qubits. 

Following these operations, the result of the addition is obtained in the Fourier domain as a superposition state. In the final step, the inverse QFT (IQFT) is applied to transform the state back to the computational basis, yielding the final result of the multiplication.

\begin{figure}[htbp]
  \centering
  \includegraphics[width=0.35\textwidth]{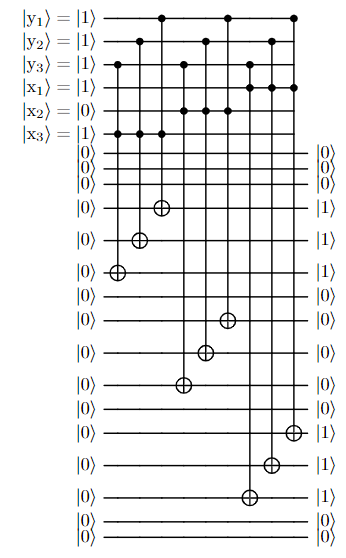}
  \caption{Three bit multiplier}
\end{figure}

\begin{figure}[htbp]
  \centering
  \includegraphics[width=0.6\textwidth]{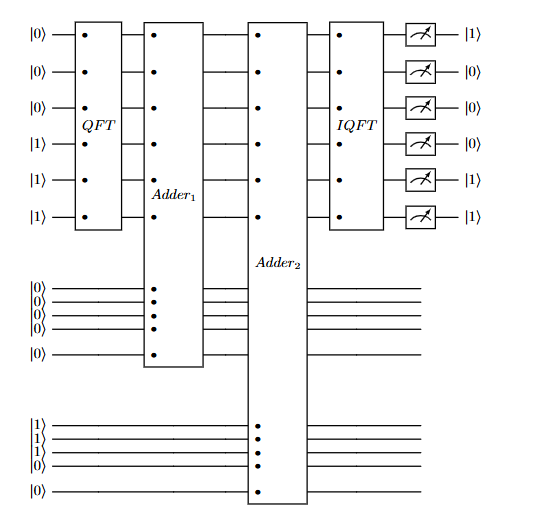}
  \caption{The quantum circuit diagam of ($\mathrm{n}$3-bit $\mathrm{N}$-input) QFT-adder.}
\end{figure}

\newpage
\section{Conclusion}

In this study, a QFT-based parallelized quantum multiplication circuit has been proposed by adapting the classical schoolbook multiplication algorithm to the quantum computing framework. The multiplication stage is implemented entirely using Toffoli gates, enabling efficient realization of bitwise multiplication operations. The intermediate results are then combined using a QFT-based parallel addition scheme.

The proposed approach significantly reduces the number of required QFT and inverse QFT operations. Instead of applying these transformations at each addition step, they are utilized only once at the beginning and once at the end of the circuit. This reduction leads to improvements in circuit depth, gate count, and overall computational efficiency.

Furthermore, a generalized quantum circuit structure has been introduced for the multiplication of two $n$-bit numbers, requiring a total of $2n^2 + n + 1$ qubits. The validity and functionality of the proposed design have been demonstrated through an example implementation for $n=3$.

Future work may focus on optimizing the circuit in terms of qubit usage and gate complexity, as well as investigating its implementation on noisy intermediate-scale quantum (NISQ) devices. Additionally, extending the proposed approach to modular multiplication and other arithmetic operations could further enhance its applicability in quantum algorithms.

\section{ACKNOWLEDGMENTS}

The authors acknowledgment support from the Scientific and Technological Research Council of Turkey (TÜBİTAK- Grant No. 122F298) 

\bibliography{references} 

@article{bennett2000,
  author  = {Bennett, Charles H. and DiVincenzo, David P.},
  title   = {Quantum Information and Computation},
  journal = {Nature},
  volume  = {404},
  number  = {6775},
  pages   = {247--255},
  year    = {2000},
  doi     = {10.1038/35005001}
}

@article{cuccaro2004,
  author  = {Cuccaro, Steven A. and Draper, Thomas G. and Kutin, Samuel A. and Moulton, David P.},
  title   = {A New Quantum Ripple-Carry Addition Circuit},
  journal = {arXiv preprint},
  year    = {2004},
  eprint  = {quant-ph/0410184},
  archivePrefix = {arXiv}
}

@article{gossett1998,
  author  = {Gossett, Phil},
  title   = {Quantum Carry-Save Arithmetic},
  journal = {arXiv preprint},
  year    = {1998},
  eprint  = {quant-ph/9808061},
  archivePrefix = {arXiv}
}

@article{vedral1995,
  author  = {Vedral, Vlatko and Barenco, Adriano and Ekert, Artur},
  title   = {Quantum Networks for Elementary Arithmetic Operations},
  journal = {Physical Review A},
  volume  = {54},
  number  = {1},
  pages   = {147--153},
  year    = {1996},
  doi     = {10.1103/PhysRevA.54.147}
}

@article{florio2004,
  author  = {Florio, G. and Picca, D.},
  title   = {Quantum Implementation of Elementary Arithmetic Operations},
  journal = {arXiv preprint},
  year    = {2004},
  eprint  = {quant-ph/0403048},
  archivePrefix = {arXiv}
}

@article{beauregard2003,
  author       = {Beauregard, Stephane and others},
  title        = {Quantum Arithmetic on Galois Fields},
  journal      = {arXiv preprint},
  year         = {2003},
  eprint       = {quant-ph/0301163},
  archivePrefix= {arXiv},
  primaryClass = {quant-ph},
  url          = {http://arxiv.org/abs/quant-ph/0301163}
}

@article{draper2000,
  author  = {Draper, Thomas G.},
  title   = {Addition on a Quantum Computer},
  journal = {arXiv preprint},
  year    = {2000},
  eprint  = {quant-ph/0008033},
  archivePrefix = {arXiv}
}

@article{ruizperez2017,
  author  = {Ruiz-Perez, Lidia and Garcia-Escartin, Juan Carlos},
  title   = {Quantum Arithmetic with the Quantum Fourier Transform},
  journal = {Quantum Information Processing},
  volume  = {16},
  number  = {6},
  pages   = {152},
  year    = {2017},
  doi     = {10.1007/s11128-017-1603-1}
}

@article{sahin2020,
  author  = {Sahin, Engin},
  title   = {Quantum Arithmetic Operations Based on Quantum Fourier Transform on Signed Integers},
  journal = {International Journal of Quantum Information},
  volume  = {18},
  number  = {6},
  pages   = {2050035},
  year    = {2020},
  doi     = {10.1142/S0219749920500355}
}

@article{wang2020,
  author  = {Wang, Yuchen and others},
  title   = {Qudits and High-Dimensional Quantum Computing},
  journal = {Frontiers in Physics},
  volume  = {8},
  pages   = {589504},
  year    = {2020},
  doi     = {10.3389/fphy.2020.589504}
}

@article{pavlidis2017,
  author  = {Pavlidis, Archimedes and Floratos, Emmanuel},
  title   = {Arithmetic Circuits for Multilevel Qudits Based on Quantum Fourier Transform},
  journal = {arXiv preprint},
  year    = {2017},
  eprint  = {1707.08834},
  archivePrefix = {arXiv}
}

@article{ramezani2023,
  author  = {Ramezani, Mehdi and others},
  title   = {Quantum Multiplication Algorithm Based on the Convolution Theorem},
  journal = {Physical Review A},
  volume  = {108},
  number  = {5},
  pages   = {052405},
  year    = {2023},
  doi     = {10.1103/PhysRevA.108.052405}
}

@article{karatsuba1962,
  author  = {Karatsuba, A. A. and Ofman, Y. P.},
  title   = {Multiplication of Multidigit Numbers by Automatic Computers},
  journal = {Doklady Akademii Nauk},
  volume  = {145},
  pages   = {293--294},
  year    = {1962}
}

@article{toom1963,
  author  = {Toom, A. L.},
  title   = {The Complexity of a Scheme of Functional Elements Realizing the Multiplication of Integers},
  journal = {Soviet Mathematics Doklady},
  volume  = {3},
  pages   = {714--716},
  year    = {1963}
}

@article{schonhage1971,
  author  = {Schonhage, Arnold},
  title   = {Schnelle Multiplikation Großer Zahlen},
  journal = {Computing},
  volume  = {7},
  pages   = {281--292},
  year    = {1971}
}

@inproceedings{furer2007,
  author  = {F{\"u}rer, Martin},
  title   = {Faster Integer Multiplication},
  booktitle = {Proceedings of the 39th Annual ACM Symposium on Theory of Computing},
  pages   = {57--66},
  year    = {2007}
}

@article{harvey2021,
  author  = {Harvey, David and van der Hoeven, Joris},
  title   = {Integer Multiplication in Time O(n log n)},
  journal = {Annals of Mathematics},
  volume  = {193},
  number  = {2},
  pages   = {563--617},
  year    = {2021}
}

@article{cakmak2023,
  author  = {Cakmak, Selcuk and others},
  title   = {Quantum Fourier Transform-Based Arithmetic Logic Unit on a Quantum Processor},
  journal = {Annalen der Physik},
  year    = {2023},
  pages   = {2300457},
  doi     = {10.1002/andp.202300457}
}

@article{cho2020,
  author  = {Cho, Seong-Min and others},
  title   = {Quantum Modular Multiplication},
  journal = {IEEE Access},
  volume  = {8},
  pages   = {213244--213252},
  year    = {2020},
  doi     = {10.1109/ACCESS.2020.3039167}
}

@article{zhan2023,
  author  = {Zhan, Junpeng},
  title   = {Quantum Multiplier Based on Exponent Adder},
  journal = {arXiv preprint},
  year    = {2023},
  eprint  = {2309.10204},
  archivePrefix = {arXiv}
}

@article{chuang2018,
  author  = {Rines, R. and Chuang, Isaac L.},
  title   = {High Performance Quantum Modular Multipliers},
  journal = {arXiv preprint},
  year    = {2018},
  eprint  = {1801.01081},
  archivePrefix = {arXiv}
}

@article{camps2021,
  author  = {Camps, Daan and others},
  title   = {Quantum Fourier Transform Revisited},
  journal = {Numerical Linear Algebra with Applications},
  volume  = {28},
  number  = {1},
  pages   = {e2331},
  year    = {2021},
  doi     = {10.1002/nla.2331}
}

@article{jozsa1998,
  author  = {Jozsa, Richard},
  title   = {Quantum Algorithms and the Fourier Transform},
  journal = {Proceedings of the Royal Society A},
  volume  = {454},
  number  = {1969},
  pages   = {323--337},
  year    = {1998},
  doi     = {10.1098/rspa.1998.0163}
}

@article{barenco1996,
  author  = {Barenco, Adriano and others},
  title   = {Approximate Quantum Fourier Transform and Decoherence},
  journal = {Physical Review A},
  volume  = {54},
  number  = {1},
  pages   = {139--146},
  year    = {1996},
  doi     = {10.1103/PhysRevA.54.139}
}

@article{cao2011,
  author  = {Cao, Ye and others},
  title   = {Quantum Fourier Transform and Phase Estimation in Qudit System},
  journal = {Communications in Theoretical Physics},
  volume  = {55},
  number  = {5},
  pages   = {790--794},
  year    = {2011},
  doi     = {10.1088/0253-6102/55/5/11}
}

@article{kurt2024,
  author  = {Kurt, Murat and Kaltehei, Ayda and Gençten, Azmi and Çakmak, Selçuk},
  title   = {Scalable quantum circuit design for QFT-based arithmetic},
  journal = {arXiv preprint},
  volume  = {},
  number  = {},
  pages   = {},
  year    = {2024},
  doi     = {},
}
\bibliographystyle{apsrev4-2} 

\end{document}